%% file: SAM_v4.tex
\documentclass[conference]{IEEEtran}
\IEEEoverridecommandlockouts

\usepackage{cite}
\usepackage{amsmath,amssymb,amsfonts}
\usepackage{graphicx}
\usepackage{textcomp}
\usepackage{xcolor}
\def\BibTeX{{\rm B\kern-.05em{\sc i\kern-.025em b}\kern-.08em
    T\kern-.1667em\lower.7ex\hbox{E}\kern-.125emX}}

\usepackage{subfig}
\usepackage{ upgreek }
\usepackage{graphicx}
\usepackage{graphics}
\usepackage{epstopdf}
\usepackage{amsmath}
\usepackage{multirow}
\usepackage{bm}
\usepackage{enumerate}
\usepackage{mathdots}
\usepackage{todonotes}
\usepackage{algorithmicx}
\usepackage{algpseudocode}
\usepackage{algorithm}
\usepackage{verbatim}
\usepackage{balance}
\usepackage{cite}
\usepackage[hyphens]{url}
\usepackage{acro}

\usepackage{color}
\definecolor{purple(x11)}{rgb}{0.63, 0.36, 0.94}
\definecolor{cadmiumgreen}{rgb}{0.0, 0.42, 0.24}

\input{Other/macros.tex}
\input{Other/acronyms.tex}

\usepackage{pifont}

\newcommand{\be}{\begin{eqnarray}}
\newcommand{\ee}{\end{eqnarray}}

\def\bx{{\mathbf{x}}}

\def\bz{{\mathbf{z}}}
\def\b0{{\mathbf{0}}}

\def\bI{{\mathbf{I}}}

\def\bsf0{{\bm{\mathsf{0}}}}

\begin{document}
\title{Joint Initial Access and Localization in Millimeter Wave Vehicular Networks: a Hybrid Model/Data Driven Approach}
\author{Yun Chen$^{\dag}$, Joan Palacios$^{\dag}$, Nuria Gonz\'{a}lez-Prelcic$^{\dag}$, Takayuki Shimizu$^\ddag$ and Hongsheng Lu$^\ddag$\thanks{ This work has been partially funded by Toyota Motor North America.} \\
	$^{\dag}$ North Carolina State University, Email: \{ychen273, jbeltra, ngprelcic\}@ncsu.edu \\
	$^\ddag$ Toyota Motor North America, Email: \{takayuki.shimizu, hongsheng.lu\}@toyota.com}

\maketitle

\begin{abstract}
High resolution compressive channel estimation provides information for vehicle localization when a hybrid mmWave MIMO system is considered. Complexity and memory requirements can, however, become a bottleneck when high accuracy localization is required. An additional challenge is the need of path order information to apply the appropriate geometric relationships between the channel path parameters and the vehicle, RSU  and scatterers position. In this paper, we propose a low complexity channel estimation strategy of the angle of departure and time difference of arrival based on multidimensional orthogonal matching pursuit. We also design a deep neural network that predicts the order of the channel paths so only the LoS and first order reflections are used for localization. Simulation results obtained with realistic vehicular channels generated by ray tracing show that sub-meter accuracy can be achieved for 50\% of the users, without resorting to perfect synchronization assumptions or unfeasible all-digital high resolution MIMO architectures.
\end{abstract}

%

\section{Introduction}
Future vehicular networks will support millimeter wave (mmWave) MIMO communication to enable use cases that require high data rate communication and precise localization \cite{Choi2016}. Channel estimation is one of the key tasks in a millimeter wave transceiver to configure the required large antenna arrays and establish the communication link between the vehicle and the road side unit (RSU). Vehicle position can be obtained as a byproduct of communication, exploiting the geometric relationships between channel parameters and the RSU and vehicle positions \cite{Shahmansoori2018}.
Combining localization based on the mmWave signal with position information provided by onboard sensors can provide the required redundancy to support use cases such as automated driving \cite{Wymeersch2017}.

Prior work on  channel estimation at mmWave for initial access and single shot localization exploits the sparse nature of the channel, both to recover the channel coefficients from the received signal \cite{lee2014exploiting, bielsa2018indoor, Venugopal2017, SWOMP2018, Coma2018, Wu2019, Zhu2019}, and to relate the channel paths to the position of the RSU, vehicle and scatterers in the environment \cite{posalg2,Shahmansoori2018,Talvitie2019, Jiang2021}.  Other approaches for vehicle localization at mmWave based on channel information focus on  position tracking \cite{Wymeersch2018, chu2021vehicle},  requiring previous state information for accurate localization. In these approaches, the order of the paths is critical to support the geometric relationships for localization with only the line-of-sight (LoS) path and the first order non-line-of-sight (NLoS) components used. 

Limitations of prior work on position estimation based on compressed channel measurements include the: a) absence of strategies to classify the reflection order of estimated channel paths; b) high complexity of state-of-the-art compressive channel estimation strategies when achieving high resolution angle and delay estimates; c) use of oversimplified models which assume all digital architectures or perfect synchronization to exploit \ac{ToA} instead of \ac{TDoA}, or neglect filtering effects at the transmitter and the receiver in the channel model.

In this paper, we propose a hybrid model/data driven strategy for position estimation. First, a low complexity compressive channel estimate is found based on multidimensional orthogonal matching pursuit (MOMP) algorithm \cite{MOMP}, enabling operation in realistic 3D scenarios, with a realistic hybrid MIMO architecture and a practical channel model that does not assume perfect synchronization or neglects filtering effects. Then, a new data driven approach based on a simple deep neural network (DNN) solves  the path classification problem. Finally, a new position estimator that can operate in LoS or non-LoS channels without assuming perfect synchronization, maps the parameters from the first order reflections and the LoS path into vehicle position. Simulation results obtained with vehicular channels generated by ray tracing show the effectiveness of this model/data driven approach to obtain high accuracy position information at reduced complexity.

\section{System Model}
We consider the uplink of a hybrid analog-digital MIMO system to support joint vehicular communication and localization at mmWave bands. 
Both the transmitter  and the receiver use a \ac{URA} of size  $N_{\rm T} = N_{\rm T}^{\rm x}\times N_{\rm T}^{\rm y}$, and $N_{\rm R} = N_{\rm R}^{\rm x}\times N_{\rm R}^{\rm y}$, respectively. The number of RF chains is denoted as  $N_{\rm T}^{\rm RF}$ for the transmitter,  and $N_{\rm R}^{\rm RF}$ for the receiver. We define directions of arrival/departure modeled as unitary vectors ${\boldsymbol \theta}, {\boldsymbol \phi}\in\mathcal{S}_2\subset\mathbb{C}^3$, such that ${\boldsymbol \theta}=[\theta^{\rm x}, \theta^{\rm y}, \theta^{\rm z}]$ and ${\boldsymbol \phi}=[\phi^{\rm x}, \phi^{\rm y}, \phi^{\rm z}]$.
The steering vector for the URA can be written as the Kronecker product of their x-y components, i.e.  ${\bf a}_{\rm R}({\boldsymbol \theta})={\bf a}_{\rm R}^{\rm x}(\theta^{\rm x})\otimes {\bf a}_{\rm R}^{\rm y}(\theta^{\rm y})$ and ${\bf a}_{\rm T}({\boldsymbol \phi})={\bf a}_{\rm T}^{\rm x}(\phi^{\rm x})\otimes {\bf a}_{\rm T}^{\rm y}(\phi^{\rm y})$, where 
${\bf a}_{\rm R-T}^{\rm x-y}\in\mathbb{C}^{N_{\rm R-T}^{x-y}}$ are defined as  $[{\bf a}_{\rm R}^{\rm x-y}(\theta^{\rm x-y})]_n=e^{-i(n-1)\pi\theta^{\rm x-y}}$ and $[{\bf a}_{\rm T}^{\rm x-y}(\phi^{\rm x-y})]_n=e^{-i(n-1)\pi\phi^{\rm x-y}}$. The dash in the categorical scripts means that the expression is valid for both categorical values.
The pulse shaping function resulting from integrating all the filtering effects at the transmitter and receiver is  $p:\mathbb{R}\rightarrow\mathbb{C}$.
To account for synchronization imperfections we define the time $t_0$ as the delay between the beginning of the transmission and the beginning of the reception.
We consider a frequency selective channel with a delay tap length $D$. We define the vectors ${\bf a}_{\rm D}(\tau-t_0)\in\mathbb{C}^{D}$ as $[{\bf a}_{\rm D}(\tau-t_0)]_d = p(T_{\rm s}(d-1)-(\tau-t_0))$.
The geometric channel model with $L$ paths is considered. Each path is characterized by its complex gain $\alpha_l$, DoA  ${\boldsymbol \theta}_l$, DoD ${\boldsymbol \phi}_l$, and delay  $\tau_l$. With these definitions, the MIMO channel matrix for the $d$-th delay tap is
\begin{equation}\label{eq:Channel}
	{\bf H}_d = \sum_{l=1}^L\alpha_l{\bf a}_{R}({\boldsymbol \theta}_l){\bf a}_{T}^{\rm H}({\boldsymbol \phi}_l)[{\bf a}_{D}^{\rm H}(\tau_l-t_0)]_d.
\end{equation}

During the initial access phase, the transmitter sends a training sequence using a set of training hybrid precoders  ${\bf F}_{m_{\rm T}}\in\mathbb{C}^{N_{\rm T}\times N_{\rm T}^{\rm RF}}$, $m_{\rm T} = 1,\ldots M_{\rm T}$, which are received at the RSU by a set of training hybrid combiners ${\bf W}_{m_{\rm R}}\in\mathbb{C}^{N_{\rm R}\times N_{\rm R}^{\rm RF}}$, $m_{\rm R} = 1,\ldots M_{\rm R}$. The different combinations of training precoders and combiners lead to a set of received training sequences which are used to sound the channel and estimate the vehicle location.
The training sequence contains $Q$ symbols ${\bf s}[q]\in\mathbb{C}^{N_{\rm T}^{\rm RF}}$ such that  $\frac{1}{Q}\sum_{q=1}^Q\|{\bf s}\|^2=1$.
The received signal corresponding to the $q$-th training symbol is 
\begin{equation}\label{eq:Measure}
	{\bf y}_{m_{\rm R}, m_{\rm T}}[q] = {\bf W}_{m_{\rm R}}^{\rm H}\sum_{d=1}^{D}{\bf H}_d{\bf F}_{m_{\rm T}}{\bf s}_{m_{\rm T}}[q-d]+{\bf W}_{m_{\rm R}}^{\rm H}{\bf n}_{m_{\rm R}, m_{\rm T}}[q].
\end{equation}
where ${\bf n}_{m_{\rm R}, m_{\rm T}}[q]$ are the gaussian  noise samples. To whiten the signal in \eqref{eq:Measure}, we consider the Cholesky decomposition of the noise correlation matrix, i.e. ${\bf L}_{m_{\rm R}}{\bf L}_{m_{\rm R}}^{\rm H} = {\bf W}_{m_{\rm R}}^{\rm H}{\bf W}_{m_{\rm R}}$. The whitened received signal is defined as $\bar{\bf y}_{m_{\rm R}, m_{\rm T}}[q] = {\bf L}_{m_{\rm R}}^{-1}{\bf y}_{m_{\rm R}, m_{\rm T}}[q]$, that can also be written as 
\begin{equation}\label{eq:Measure_white}
	\bar{\bf y}_{m_{\rm R}, m_{\rm T}}[q] = \bar{\bf W}_{m_{\rm R}}^{\rm H}\sum_{d=1}^{D}{\bf H}_d{\bf F}_{m_{\rm T}}{\bf s}_{m_{\rm T}}[q-d]+\bar{\bf n}_{m_{\rm R}, m_{\rm T}}[q],
\end{equation}
where $\bar{\bf n}_{m_{\rm R}, m_{\rm T}}[q] = {\bf L}_{m_{\rm R}}^{-1}{\bf W}_{m_{\rm R}}^{\rm H}{\bf n}_{m_{\rm R}, m_{\rm T}}[q]$ and $\bar{\bf W}_{m_{\rm R}}^{\rm H} = {\bf L}_{m_{\rm R}}^{-1}{\bf W}_{m_{\rm R}}^{\rm H}$.
The final observation matrix for a given combination of training precoder and combiner and its corresponding noise are defined by stacking the received symbols and noise samples as \begin{multline}\label{eq:Symbol2signal}
	{\bf Y}_{m_{\rm R}, m_{\rm T}} = [\bar{\bf y}_{m_{\rm R}, m_{\rm T}}[1], \cdots, \bar{\bf y}_{m_{\rm R}, m_{\rm T}}[Q]],\\
	{\bf N}_{m_{\rm R}, m_{\rm T}} = [\bar{\bf n}_{m_{\rm R}, m_{\rm T}}[1], \cdots, \bar{\bf n}_{m_{\rm R}, m_{\rm T}}[Q]].
\end{multline}

\section{Joint channel estimation and localization system}


\subsection{MOMP-based channel estimation}
The MOMP problem \cite{MOMP} consists of finding a sparse tensor given an observation ${\bf Y}\in\mathbb{C}^{N_{\rm q}\times N_{\rm o}}$, a collection of $N_{\rm D}$ sparsifying dictionaries ${\bf \Psi}_{k}\in\mathbb{C}^{N_k^{\rm s}\times N_k^{\rm a}}$,  and a measurement tensor ${\bf \Phi}\in\mathbb{C}^{N_q\times\otimes_{k=1}^{N_{\rm D}}N_k^{\rm s}}$. The tensor that contains the  sparse coefficients is denoted as ${\bf C}\in\mathbb{C}^{\otimes_{k=1}^{N_{\rm D}}N_k^{\rm s}\times N_{\rm o}}$.
To ease the notation, the set entry coordinate combinations and the set of dictionary index combinations are defined to cycle over each dictionary atom entry index and dictionary atom index like $\mathcal{I} = \{{\bf i}=(i_1, \ldots, i_{N_{\rm D}})\in\mathbb{N}^{N_{\rm D}}\text{ s.t. }i_k \leq N_k^{\rm s}\quad\forall k \leq N_{\rm D}\}$ and $\mathcal{J} = \{{\bf j}=(j_1, \ldots, j_{N_{\rm D}})\in\mathbb{N}^{N_{\rm D}}\text{ s.t. }j_{d} \leq N_k^{\rm a}\quad\forall k \leq N_{\rm D}\}$ respectively.
The coefficients support is defined as $\mathcal{C}\in\mathcal{J}$ such that ${\bf j}\in\mathcal{C}$ if and only if $\|[{\bf C}]_{{\bf j}, :}\| > 0$ and must satisfy the sparsity condition $|\mathcal{C}|\leq N_{\rm p}$. The MOMP problem can then be formulated as
\begin{equation}\label{eq:MOMP}
	\min_{\bf C}\left(\sum_{i_{\rm m} = 1}^{N^{\rm m}}\left\|{\bf Y}-\sum_{{\bf i}\in\mathcal{I}}\sum_{{\bf j}\in\mathcal{J}}[{\bf \Phi}]_{:, {\bf i}}\left(\prod_{k = 1}^{N_{\rm D}}[{\bf \Psi}_{k}]_{i_k, j_k}\right)[{\bf C}]_{{\bf j}, :}\right\|^2\right).
\end{equation}

This approach enables sparse recovery with lower complexity, since projections over  a set of dictionaries instead of a larger single dictionary are considered. To reduce memory requirements, which become a bottleneck when operating with large arrays, we focus on the estimation of delays, DoDs and equivalent gains, which include the effect of the DoAs and the path complex gains. 
Considering discretized domains for the x and y components of the DoD, $\{\bar{\phi}_1^{\rm x}, \ldots, \bar{\phi}_{N_1^{\rm a}}^{\rm x}\}$ and $\{\bar{\phi}_1^{\rm y}, \ldots, \bar{\phi}_{N_2^{\rm a}}^{\rm y}\}$ respectively, and also for the delay, $\{\bar{\tau}_1, \ldots, \bar{\tau}_{N_2^{\rm a}}\}$, the  sparsifying dictionaries are the ones defined by the transmit array steering vectors evaluated on the grid for the DoD, and the pulse shaping function evaluated on the grid for the delay. Thus, 
\begin{multline}\label{eq:Dictionaries}
	{\bf \Psi}_1 = [{\bf a}_{\rm T}^{\rm x}(\bar{\phi}_1^{\rm x})^*, \cdots, {\bf a}_{\rm T}^{\rm x}(\bar{\phi}_{N_1^{\rm a}}^{\rm x})^*]\\
	{\bf \Psi}_2 = [{\bf a}_{\rm T}^{\rm y}(\bar{\phi}_1^{\rm y})^*, \cdots, {\bf a}_{\rm T}^{\rm y}(\bar{\phi}_{N_2^{\rm a}}^{\rm y})^*]\\
	{\bf \Psi}_3 = [{\bf a}_{\rm D}(\bar{\tau}_1), \cdots, {\bf a}_{\rm D}(\bar{\tau}_{N_3^{\rm a}})].
\end{multline}
To obtain a compact expression for the sparse tensor  ${\bf C}$, we define  the vector ${\boldsymbol \beta}_l\in\mathbb{C}^{N_{\rm R}^{\rm RF}M_{\rm R}}$ containing the complex gain and angular response contribution as
\begin{equation}\label{eq:Beta}
	[{\boldsymbol \beta}_l]_{N_{\rm R}^{\rm RF}m_{\rm R}+n_{\rm R}^{\rm RF}}=\alpha_l[{\bf W}_{m_{\rm R}}]^{\rm H}_{:, n_{\rm R}^{\rm RF}}{\bf a}_{\rm R}({\boldsymbol \theta}_l).
\end{equation}
By ignoring quantization effects, ${\bf C}$ can be written 
as
\begin{equation}\label{eq:Coefficients}
	[{\bf C}]_{{\bf j}, :} = \left\lbrace\begin{array}{cl}
		{\boldsymbol \beta}_l & \text{if }\begin{array}{c}
			\phi_l^{\rm x} = \bar{\phi}_{j_1}^{\rm x}, \phi_l^{\rm y} = \bar{\phi}_{j_2}^{\rm y}\\
			\tau_l-t_0 = \bar{\tau}_{j_3}
		\end{array}\\
		0 & \text{otherwise}
	\end{array}\right..
\end{equation}
The measurement matrix is defined as 
\begin{equation}
	[\Phi]_{Qm_{\rm T}+q, {\bf i}} = [{\bf F}_{m_{\rm T}}{\bf s}[q-i_3]]_{i_1N_{\rm T}^{\rm x}+i_2}.
\end{equation}
The whole observation matrix is built from \eqref{eq:Symbol2signal} as
\begin{align}
	{\bf Y} = \left[\begin{array}{ccc}
		{\bf Y}_{1, 1}^{\rm T} & \cdots & {\bf Y}_{M_{\rm R}, 1}^{\rm T}\\
		\vdots & \ddots & \vdots\\
		{\bf Y}_{1, M_{\rm T}}^{\rm T} & \cdots & {\bf Y}_{M_{\rm R}, M_{\rm T}}^{\rm T}\\
	\end{array}\right].
\end{align}
Analogously, the noise matrix {\bf N} can also be constructed from the noise components in \eqref{eq:Symbol2signal}.
Using these definitions, \eqref{eq:Measure_white} can be rewritten as
\begin{equation}\label{eq:FullMeasurement}
	{\bf Y}=\sum_{{\bf i}\in\mathcal{I}}\sum_{{\bf j}\in\mathcal{J}}[{\bf \Phi}]_{:, {\bf i}}\left(\prod_{k = 1}^{N_{\rm D}}[{\bf \Psi}_{k}]_{i_k, j_k}\right)[{\bf C}]_{{\bf j}, :}+{\bf N}.
\end{equation}
Since ${\bf N}$ can be modeled as white noise, the maximum likelihood estimator estimator for ${\bf C}$ can be obtained by solving \eqref{eq:MOMP}. A suitable approach to solve this problem is the MOMP algorithm described in \cite{MOMP}.

\subsection{DoA Retrieval}
From the output of the MOMP algorithm,  the DoA contribution embedded in ${\bf C}$ can also be obtained. To this aim, we group the multiple training combiners into a single variable ${\bf W} = [\bar{\bf W}_1, \cdots, \bar{\bf W}_{M_{\rm R}}]$.
This way, ${\boldsymbol \beta}_l$ in \eqref{eq:Beta} can be written as ${\boldsymbol \beta}_l = \alpha_l{\bf W}^{\rm H}{\bf a}_{\rm R}({\boldsymbol \theta}_l)$.
From this equation, and using the estimated $\hat{\boldsymbol \beta}_l$, we can obtain an estimation of the direction of arrival as 
\begin{equation}
	\hat{\boldsymbol \theta}_l = \arg\max_{{\boldsymbol \theta}} \hat{\boldsymbol \beta}_l^{\rm H}{\bf W}^\dagger{\bf a}_{\rm R}({\boldsymbol \theta}).
\end{equation}
with ${\bf W}^\dagger$ the pseudo-inverse of ${\bf W}$.  
We solve this maximum projection problem by discretizing the problem and evaluating the multiple discrete values of ${\boldsymbol \theta}$.

\subsection{Localization based on path geometry}
\begin{figure}[h!]
	\centering
	\subfloat[LOS+NLOS localization]{%
		\label{LOSNLOS_loc_model}
		\includegraphics[width=0.24\textwidth]{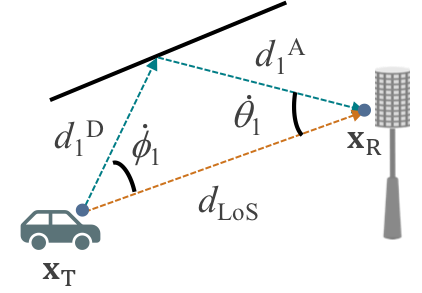}}
	\subfloat[NLOS only localization]{%
		\label{NLOSOnly_loc_model}
		\includegraphics[width=0.24\textwidth]{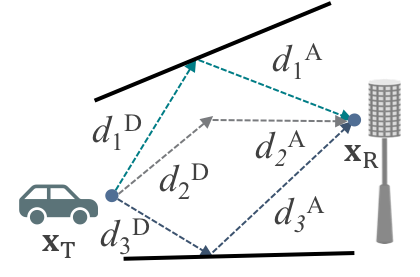}}
	\caption{Localization using: (a) LOS+NLOS paths; (b) NLOS only.}
	\label{loc_model}
\end{figure}

We will exploit the geometric relationships illustrated in Fig. \ref{loc_model} for LoS and NLoS channels to convert the channel parameters into an estimate of the vehicle position. Note that these relationships apply to first order reflections only. Higher orders reflections will be identified using the  DNN proposed in Section \ref{path_order_got} and will be discarded   for localization. 

\vspace*{-3mm}
For the LoS+NLoS case shown in Fig.\ref{LOSNLOS_loc_model}, we can compute the angle that the NLOS path creates with the LoS path using the expressions $\dot{\theta}_l=\arccos({\boldsymbol \theta}_{\rm LoS}^{\rm T}{\boldsymbol \theta}_l)$ and $\dot{\phi}_l=\arccos({\boldsymbol \phi}_{\rm LoS}^{\rm T}{\boldsymbol \phi}_l)$.
Considering $d_l^{\rm A}$ and $d_l^{\rm D}$ to be the distance between the receiver/transmitter and the reflection point, and $d_{\rm LoS}$ the distance between receiver and transmitter, the sine theorem applied to the triangle formed by the location of the receiver, transmitter and reflection point fulfills
\begin{equation}
\frac{d_{\rm LoS}}{\sin(\dot{\theta}_l+\dot{\phi}_l)}=\frac{d_l^{\rm D}}{\sin(\dot{\theta}_l)}=\frac{d_l^{\rm A}}{\sin(\dot{\theta}_l)}.
\end{equation}
By defining the known parameter $\Delta d_l=d_l^{\rm A} + d_l^{\rm D} - d_{\rm LoS} = c\left((\tau_l-t_0)-(\tau_1-t_0)\right)$, the distance $d_{\rm LOS}$ is given by
\begin{equation}
d_{\rm LoS} =\frac{\Delta d_l \sin(\dot{\theta}_l + \dot{\phi}_l)}{\sin(\dot{\theta}_l)+ \sin(\dot{\phi}_l)-\sin(\dot{\theta}_l + \dot{\phi}_l)}.
\end{equation}
With $N_{\rm p}$ 1st-order NLoS paths available, we can group the different variables into vectors $\dot{\boldsymbol \theta}=[\dot{\theta}_1, \cdots, \dot{\theta}_{N_{\rm p}}]$,  $\dot{\boldsymbol \phi}=[\dot{\phi}_1, \cdots, \dot{\phi}_{N_{\rm p}}]$ and  $\Delta {\bf d}=[\Delta d_1, \cdots, \Delta d_{N_{\rm p}}]$, and the distance between the vehicle and the RSU could be estimated using LS as
\begin{equation}
\hat{d}_{\rm LoS} =\frac{<\Delta {\bf d}_l \sin(\dot{\boldsymbol \theta} + \dot{\boldsymbol \phi}), \sin(\dot{\boldsymbol \theta})+ \sin(\dot{\boldsymbol \phi})-\sin(\dot{\boldsymbol \theta} + \dot{\boldsymbol \phi})>}{\|\sin(\dot{\boldsymbol \theta})+ \sin(\dot{\boldsymbol \phi})-\sin(\dot{\boldsymbol \theta} + \dot{\boldsymbol \phi})\|^2}.
\end{equation}
The estimated 3D vehicle location $\hat{\bf x}_{\rm T}$ can be obtained from the 3D RSU location $\bx_{\rmR}$ as
\begin{equation}
	\hat{\bf x}_{\rm T} = {\bf x}_{\rm R} +\hat{d}_{\rm LoS}{\boldsymbol \theta}_{\rm LoS}.
\end{equation}

Regarding NLoS channels, for each independent 1st-order path $l$, the geometric relationship illustrated in Fig. \ref{NLOSOnly_loc_model} becomes
\begin{equation}\label{NLoS_loc_equ}
\begin{cases}
{\bf x}_{\rm T} = {\bf x}_{\rm R}+{\boldsymbol \theta}_l d_l^{\rmA}-{\boldsymbol \phi}_l d_l^{\rmD}\\
d_l^{\rm A}+d_l^{\rmD}=\Delta d_l+d_{\rm LoS}
\end{cases}.
\end{equation}
The expression in \eqref{NLoS_loc_equ} is a linear system of 4 equations (3+1) and 6 variables (3 corresponding to ${\bf x}_{\rm R}$, $d_{\rm LoS}$, $d_l^{\rm A}$ and $d_l^{\rm D}$).
We solve ${\boldsymbol \theta}_l$ and ${\boldsymbol \phi}_l$ and define $\Theta_l=\frac{({\boldsymbol \theta}_l+{\boldsymbol \phi}_l)({\boldsymbol \theta}_l+{\boldsymbol \phi}_l)^{\rmT}}{||{\boldsymbol \theta}_l+{\boldsymbol \phi}_l||^2}$, to get
\begin{equation}\label{eq:loc}
	(\bI-\Theta_l)({\bf x}_{\rm R}-{\boldsymbol \phi}_l\Delta d_l)=(\bI -\Theta_l)[\bI,{\boldsymbol \phi}_l][{\bf x}_{\rm T};d_{\rm LoS}].
\end{equation}
When $N_p\geq 3$ 1st-order reflections exist, $[\hat{\bx}_{\rmT};\hat{d}_{\rm LoS}]$ can be obtained from \eqref{eq:loc} using LS estimation as
\begin{equation}\label{eq:final_loc}
[{\bf x}_{\rm T};d_{\rm LoS}] = {\bf A}^{-1}{\bf b},
\end{equation}
for ${\bf A} = \sum_{l=1}^{N_{\rm p}}[\bI,{\boldsymbol \phi}_l]^{\rm T}(\bI -\Theta_l)[\bI,{\boldsymbol \phi}_l]$ and ${\bf b} = \sum_{l=1}^{N_{\rm p}}[\bI,{\boldsymbol \phi}_l]^{\rm T}(\bI-\Theta_l)({\bf x}_{\rm R}-{\boldsymbol \phi}_l\Delta d_l)$.

\subsection{DNN-based Channel Path Order Classification}\label{path_order_got}
\begin{figure}[h!]
	\centering
	\includegraphics[width=\columnwidth]{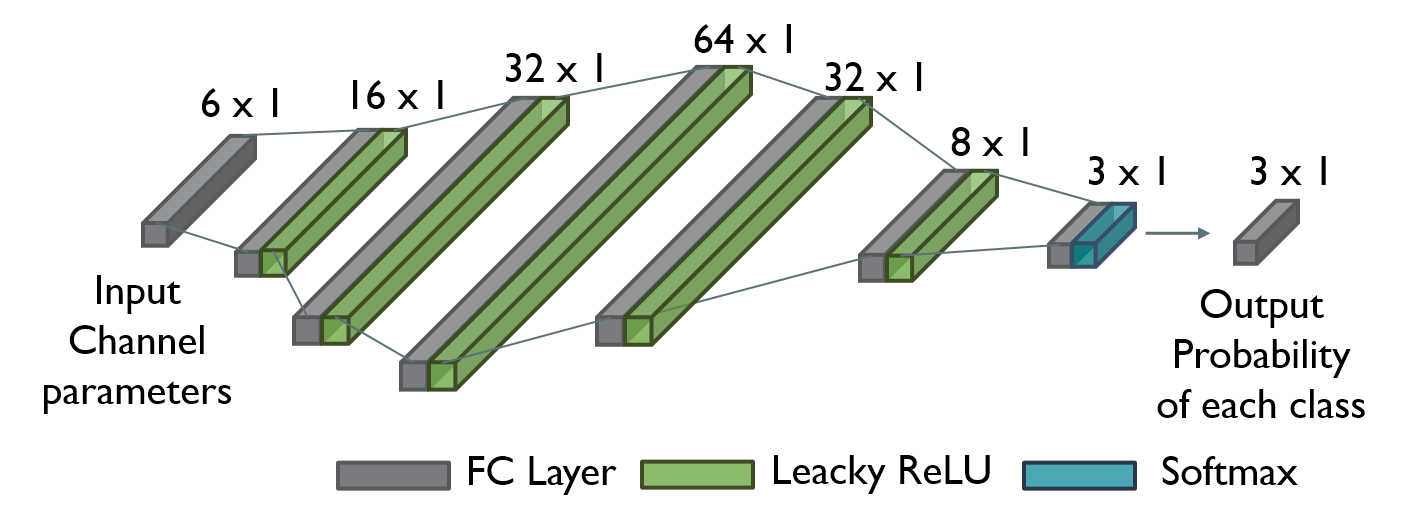}
	\caption{Illustration of the path order classification network.}
	\label{classify_DNN_arch}
\end{figure}

The path parameters to be input into the DNN for path classification are organized in vector form as  $\bz=[|\alpha|^2$, $\tau$, $\theta_{\rm az}$, $\theta_{\rm el}$, $\phi_{\rm az}$, $\phi_{\rm el}]$, where the path power gain can be extracted as $|\alpha|^2=|{\boldsymbol \beta}_l^{\rm H}{\bf W}^\dagger{\bf a}_{\rm R}({\boldsymbol \theta})|^2$, and the DoA and DoD have been transformed into polar coordinates.
Since only the LoS (if any) and 1st-order reflections are required for vehicle localization, the classification categories are defined to be: 1) LoS path; 2)1st-order NLoS path; and 3) other. For the input  $\bz$, the desired network output is defined as
\begin{equation}
g({\bf z})=\left\lbrace
\begin{array}{ccl}
{[1, 0, 0]}^{\rm T} & \text{if} & \text{LoS}\\
{[0, 1, 0]}^{\rm T} & \text{if} & \text{1st order}\\
{[0, 0, 1]}^{\rm T} & \text{if} & \text{other}
\end{array}
\right..
\end{equation}
We fit $g$ with a neural network $\hat{g}$ and define the path classification $c({\bf z})=\arg\max g({\bf z})$ and $\hat{c}({\bf z})=\arg\max \hat{g}({\bf z})$ that outputs the index of the highest entry of the output and thus the category or estimated category.
The proposed network architecture is shown in Fig. \ref{classify_DNN_arch}. 

\vspace*{-2mm}
As classifying higher order paths as LoS or first order reflections impacts localization error much more than miss-classifying a LoS path or first order reflection as a higher order path, the weighted cross entropy loss is adopted instead of traditional cross entropy loss for network training, i.e., 
\begin{equation}
\mathcal{L}({\bf z})=-e^{-\eta (c({\bf z})-\hat{c}({\bf z}))}\log(g({\bf z})^{\rm T}\hat{g}({\bf z})),
\end{equation}
where $\eta$ is the customized weight coefficient.

%

\section{Results}
We acquire the channel dataset via 500 ray-tracing simulations in Rosslyn city, Virginia.
For every simulation, vehicles are randomly distributed on the 4 lanes, and 4 of them are active and communicate with the \ac{BS}. Each vehicle is equipped with 4 \ac{URA}s on the top, and the RSU has one \ac{URA} facing the road. The parameters regarding the vehicle and the urban environment settings follow the deployment in \cite{ali2020passive}. 

For the communication system, two sets of antenna settings are used: (1) $N_{\rmT} = N_{\rmT}^{\rmx}\times N_{\rmT}^{\rmy} = 4\times4$ and $N_{\rmR} = N_{\rmR}^{\rmx}\times N_{\rmR}^{\rmy} = 8\times 8$; (2) $N_{\rmT} = N_{\rmT}^{\rmx}\times N_{\rmT}^{\rmy} = 8\times 8$ and $N_{\rmR} = N_{\rmR}^{\rmx}\times N_{\rmR}^{\rmy} = 16\times 16$. The numbers of RF chains for the \ac{TX} and \ac{RX} are $N_{\rmT}^{\rm RF}=2$ and $N_{\rmR}^{\rm RF}=4$ with antenna setting (1), and $N_{\rmT}^{\rm RF}=4$ and $N_{\rmR}^{\rm RF}=8$ with setting (2). The communication system operates at a carrier frequency $f_c=73$ GHz with a bandwidth $B_c=1$ GHz. The transmit power is configured at $P_t=20$ dBm and $P_t=40$ dBm. The raised-cosine filter is used as pulse shaping function. The sampling frequency is $f_s=1.76$ GHz, and the number of time-domain taps is fixed to $N_c=64$. 

During the channel estimation process, the codebooks and dictionaries are designed  as in \cite{MOMP}. The pilot signals are extracted from the first row of a $64\times 64$ Hadamard matrix with  zero-padding before and after the pilots. With antenna setting (1) 128 training frames are sent, while for antenna setting (2) 512 frames are considered.

The channel dataset is randomly divided by 3:1 to form the training and validation datasets for path order classification. The learning rate is 0.001 with a decay of 0.95 per 200 epochs. The total number of training epochs is set to 1000, with an early stopping based on the convergence of validation loss. The parameter $\eta$ has been set to 0.2. The  overall classification accuracy achieves 99.28\%. The accuracy for each independent class is 97.88\%, 99.21\%, and 98.91\%.

\begin{figure}[h!]
	\centering
	\includegraphics[width=.7 \columnwidth]{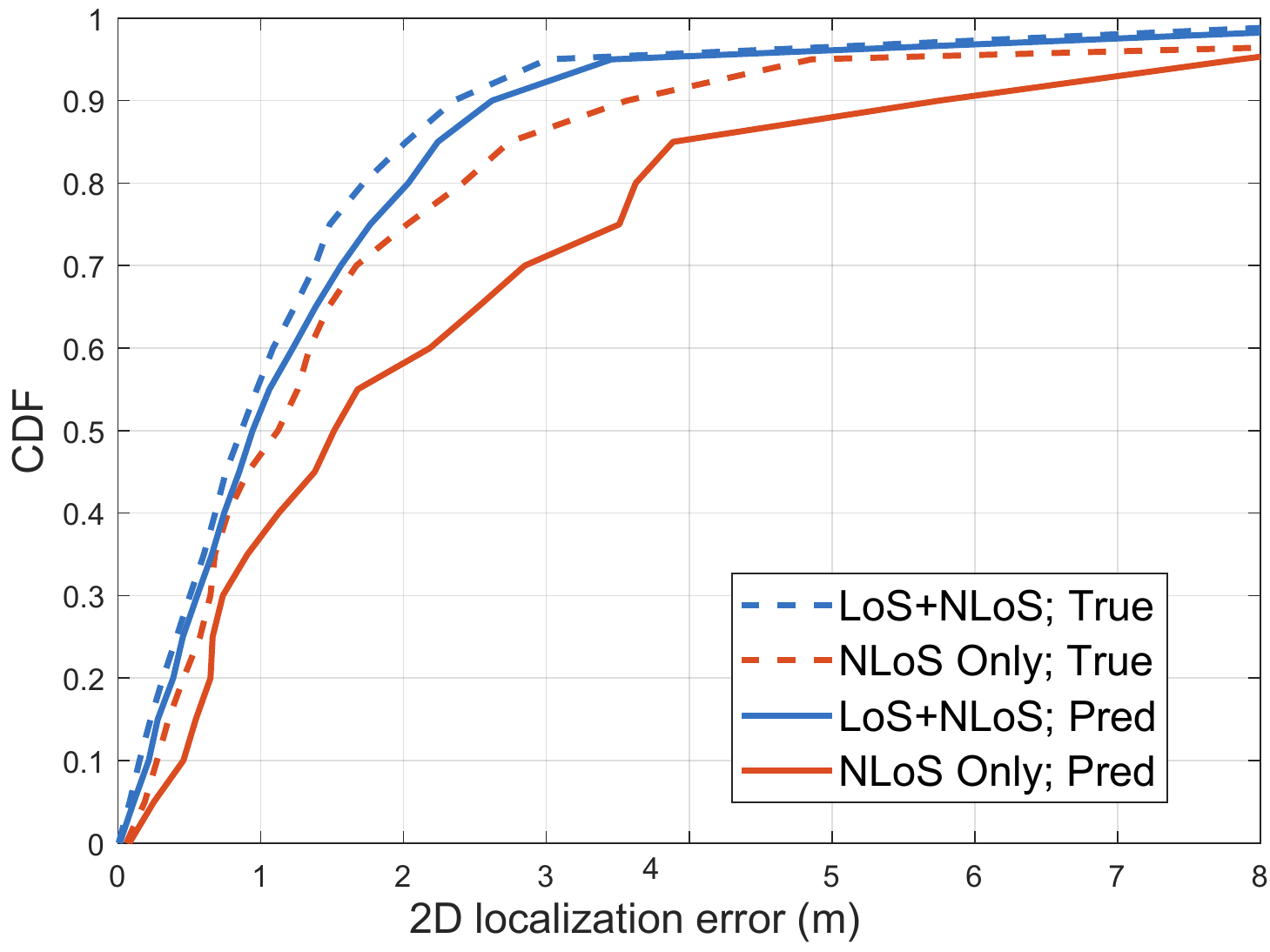}
	\caption{Comparison of the 2D localization error using the true and predicted path order based on estimated channels for $P_t=40$ dBm and antenna setting (2).}
	\label{Comparison_TrueOrPred_Inter}
\end{figure}

To understand the impact of path classification errors in the localization performance we compute the cumulative distribution function (cdf) of the  localization error when the true and predicted path orders are considered. Fig. \ref{Comparison_TrueOrPred_Inter} shows this cdf both for channels with and without a LoS component. The path order classsification error has a small impact on channel with a LoS component, but localization for NLoS channels is quite sensitive to errors in the path order.   

\vspace*{-2mm}
The localization error using different system parameters settings is summarized in Table \ref{loc_perform_table} for 5\%, 50\%, 80\% and 90\%  of the best users when using the true path order and the predicted path order in the localization algorithm.  Sub-meter accuracy is achieved for 50\% of the users in channels with a LoS components and antenna setting 2, while for 80\% of the users only accuracies below 2 m can be guaranteed. Performance degrades for antenna setting (1) and when NLoS channels are considered.
\begin{table}[h!]
	\centering
	\resizebox{\columnwidth}{!}{
		\begin{tabular}{|cc|cccc|cccc|}
			\hline
			\multicolumn{2}{|c|}{\multirow{2}{*}{Settings}}                                                            & \multicolumn{4}{c|}{LoS+NLoS}                                                          & \multicolumn{4}{c|}{NLoS Only}                                                           \\ \cline{3-10} 
			\multicolumn{2}{|c|}{}                                                                                     & \multicolumn{1}{c|}{5th}   & \multicolumn{1}{c|}{50th}  & \multicolumn{1}{c|}{80th}  & 95th  & \multicolumn{1}{c|}{5th}   & \multicolumn{1}{c|}{50th}  & \multicolumn{1}{c|}{80th}  & 95th  \\ \hline
			\multicolumn{1}{|c|}{\multirow{2}{*}{\begin{tabular}[c]{@{}c@{}}4x4\&8x8; \\ 20dBm\end{tabular}}}   & True & \multicolumn{1}{c|}{0.104} & \multicolumn{1}{c|}{1.646} & \multicolumn{1}{c|}{3.514} & 5.970 & \multicolumn{1}{c|}{0.186} & \multicolumn{1}{c|}{2.313} & \multicolumn{1}{c|}{3.930} & 7.702 \\ \cline{2-10} 
			\multicolumn{1}{|c|}{}                                                                              & Pred & \multicolumn{1}{c|}{0.133} & \multicolumn{1}{c|}{2.061} & \multicolumn{1}{c|}{4.335} & 6.785 & \multicolumn{1}{c|}{0.332} & \multicolumn{1}{c|}{2.854} & \multicolumn{1}{c|}{4.632} & 7.586 \\ \hline
			\multicolumn{1}{|c|}{\multirow{2}{*}{\begin{tabular}[c]{@{}c@{}}4x4\&8x8; \\ 40dBm\end{tabular}}}   & True & \multicolumn{1}{c|}{0.102} & \multicolumn{1}{c|}{1.618} & \multicolumn{1}{c|}{3.572} & 6.009 & \multicolumn{1}{c|}{0.225} & \multicolumn{1}{c|}{2.661} & \multicolumn{1}{c|}{4.405} & 7.214 \\ \cline{2-10} 
			\multicolumn{1}{|c|}{}                                                                              & pred & \multicolumn{1}{c|}{0.153} & \multicolumn{1}{c|}{1.785} & \multicolumn{1}{c|}{4.153} & 6.626 & \multicolumn{1}{c|}{0.315} & \multicolumn{1}{c|}{3.298} & \multicolumn{1}{c|}{5.241} & 7.886 \\ \hline
			\multicolumn{1}{|c|}{\multirow{2}{*}{\begin{tabular}[c]{@{}c@{}}8x8\&16x16; \\ 20dBm\end{tabular}}} & True & \multicolumn{1}{c|}{0.070} & \multicolumn{1}{c|}{0.936} & \multicolumn{1}{c|}{1.797} & 3.114 & \multicolumn{1}{c|}{0.178} & \multicolumn{1}{c|}{1.288} & \multicolumn{1}{c|}{2.735} & 6.287 \\ \cline{2-10} 
			\multicolumn{1}{|c|}{}                                                                              & Pred & \multicolumn{1}{c|}{0.087} & \multicolumn{1}{c|}{0.961} & \multicolumn{1}{c|}{2.068} & 3.520 & \multicolumn{1}{c|}{0.369} & \multicolumn{1}{c|}{1.947} & \multicolumn{1}{c|}{4.300} & 8.481 \\ \hline
			\multicolumn{1}{|c|}{\multirow{2}{*}{\begin{tabular}[c]{@{}c@{}}8x8\&16x16; \\ 40dBm\end{tabular}}} & True & \multicolumn{1}{c|}{0.083} & \multicolumn{1}{c|}{0.865} & \multicolumn{1}{c|}{1.719} & 3.008 & \multicolumn{1}{c|}{0.189} & \multicolumn{1}{c|}{1.122} & \multicolumn{1}{c|}{2.416} & 4.856 \\ \cline{2-10} 
			\multicolumn{1}{|c|}{}                                                                              & Pred & \multicolumn{1}{c|}{0.111} & \multicolumn{1}{c|}{0.944} & \multicolumn{1}{c|}{2.044} & 3.456 & \multicolumn{1}{c|}{0.253} & \multicolumn{1}{c|}{1.514} & \multicolumn{1}{c|}{3.626} & 6.856 \\ \hline
		\end{tabular}
	}
	\caption{2D localization errors [m] for the 5th, 50th, 80th, and 95th percentiles under different antenna and power settings.}
	\label{loc_perform_table}
\end{table}

\vspace*{-3mm}
\section{Conclusions}
We developed a sparse channel estimation and localization strategy in the context of a vehicular mmWave MIMO system based  on MOMP.
High resolution estimation of the DoDs and TDoAs is directly obtained at reduced complexity, and memory requirements, while DoA information is later extracted from the equivalent complex gains of the sparse channel tensor. We also proposed a DNN that processes the estimated paths and assigns a path order. Geometric relationships that exploit all the estimated parameters and path orders were finally obtained to estimate the vehicle position  in both LoS and NLoS channels.


\bibliographystyle{IEEEtran}
\bibliography{refs,SAM_ref}

\end{document}

%% file: Other/macros.tex
\usepackage{amsmath,amssymb}

\newcommand{\bx}{{\mathbf{x}}}

\newcommand{\bz}{{\mathbf{z}}}

\newcommand{\bI}{{\mathbf{I}}}

\newcommand{\rmx}{{\mathrm{x}}}
\newcommand{\rmy}{{\mathrm{y}}}

\newcommand{\rmA}{{\mathrm{A}}}

\newcommand{\rmD}{{\mathrm{D}}}

\newcommand{\rmR}{{\mathrm{R}}}

\newcommand{\rmT}{{\mathrm{T}}}

\makeatletter
\def\munderbar#1{\underline{\sbox\tw@{$#1$}\dp\tw@\z@\box\tw@}}
\makeatother

%% file: Other/acronyms.tex
\DeclareAcronym{3GPP}{
  short=3GPP,
  long=3rd generation partnership project
}
\DeclareAcronym{5G}{
  short=5G,
  long=fifth-generation
}
\DeclareAcronym{ADC}{
  short=ADC,
  long=analog-to-digital converter
}
\DeclareAcronym{AMP}{
  short=AMP,
  long=approximate message passing
}
\DeclareAcronym{AoA}{
  short=AoA,
  long=angle-of-arrival
}
\DeclareAcronym{AoD}{
  short=AoD,
  long=angle-of-departure
}
\DeclareAcronym{APS}{
  short=APS,
  long=azimuth power spectrum
}
\DeclareAcronym{AV}{
  short=AV,
  long=autonomous vehicle
}
\DeclareAcronym{BS}{
  short=BS,
  long=base station
}
\DeclareAcronym{BSM}{
  short=BSM,
  long=basic safety message
}
\DeclareAcronym{CDF}{
	short=CDF,
	long=cumulative distribution function
}
\DeclareAcronym{CP}{
  short=CP,
  long=cyclic-prefix
}
\DeclareAcronym{DFT}{
  short=DFT,
  long=discrete Fourier transform
}
\DeclareAcronym{DL}{
  short=DL,
  long=downlink
}
\DeclareAcronym{DNN}{
  short=DNN,
  long=deep neural network
}
\DeclareAcronym{DSRC}{
  short=DSRC,
  long=dedicated short-range communication
}
\DeclareAcronym{FC}{
  short=FC,
  long=fully connected
}
\DeclareAcronym{FDD}{
  short=FDD,
  long=frequency division duplex
}
\DeclareAcronym{FMCW}{
  short=FMCW,
  long=frequency modulated continuous wave
}
\DeclareAcronym{FoV}{
  short=FoV,
  long=field-of-view
}
\DeclareAcronym{GNSS}{
  short=GNSS,
  long=global navigation satellite system
}
\DeclareAcronym{GPS}{
  short=GPS,
  long=global positioning system
}
\DeclareAcronym{LIDAR}{
  short=LIDAR,
  long=Light detection and ranging
}
\DeclareAcronym{LOS}{
  short=LOS,
  long=line-of-sight
}
\DeclareAcronym{LPF}{
  short=LPF,
  long=low pass filter
}
\DeclareAcronym{LTE}{
  short=LTE,
  long=long term evolution
}
\DeclareAcronym{LS}{
	short=LS,
	long=least square
}
\DeclareAcronym{MIMO}{
  short=MIMO,
  long=multiple-input multiple-output
}
\DeclareAcronym{mmWave}{
  short=mmWave,
  long=millimeter-wave
}
\DeclareAcronym{MRR}{
  short=MRR,
  long=medium range radar
}
\DeclareAcronym{NLOS}{
  short=NLOS,
  long=non-line-of-sight
}
\DeclareAcronym{NR}{
  short=NR,
  long=new radio
}
\DeclareAcronym{OFDM}{
  short=OFDM,
  long=orthogonal frequency-division multiplexing
}
\DeclareAcronym{ppm}{
  short=ppm,
  long=parts-per-million
}
\DeclareAcronym{RF}{
	short=RF,
	long=radio frequency
}

\DeclareAcronym{RMS}{
  short=RMS,
  long=root-mean-square
}
\DeclareAcronym{RPE}{
  short=RPE,
  long=relative precoding efficiency
}
\DeclareAcronym{RSU}{
  short=RSU,
  long=roadside unit
}

\DeclareAcronym{RX}{
  short=RX,
  long=receiver
}
\DeclareAcronym{SNR}{
  short=SNR,
  long=signal-to-noise ratio
}
\DeclareAcronym{TDoA}{
  short=TDoA,
  long=time difference of arrival
}

\DeclareAcronym{ToA}{
  short=ToA,
  long=time of arrival
}

\DeclareAcronym{TX}{
  short=TX,
  long=transmitter
}

\DeclareAcronym{UL}{
  short=UL,
  long=uplink
}

\DeclareAcronym{ULA}{
  short=ULA,
  long=uniform linear array
}
\DeclareAcronym{URA}{
	short=URA,
	long=uniform rectangular array
}

\DeclareAcronym{V2I}{
  short=V2I,
  long=vehicle-to-infrastructure
}
\DeclareAcronym{V2V}{
  short=V2V,
  long=vehicle-to-vehicle
}
\DeclareAcronym{V2X}{
  short=V2X,
  long=vehicle-to-everything
}
\DeclareAcronym{VRU}{
  short=VRU,
  long=vulnerable road user
}